\documentclass[a4paper,11pt]{article}
\usepackage{pos}

\usepackage{tikz}
\usepackage{amsmath}

\title{How well does NRQCD describe quarkonium production?}

\author*{Mathias Butenschoen}

\affiliation{II. Institut f\"ur Theoretische Physik, Universit\"at Hamburg, \\
Luruper Chaussee 149, 22761 Hamburg, Germany} 

\emailAdd{mathias.butenschoen@desy.de}

\abstract{The question how well nonrelativistic QCD (NRQCD) factorization can describe quarkonium production has been subject to debate since its invention. We review our recent reanalysis of a classic next-to-leading order color octet (CO) long distance matrix element (LDME) fit to large transverse momentum $p_T$ $J/\psi$ and $\eta_c$ LHC production data. Our analysis differs from previous analyses of this kind not only by implementing for the first time a systematic treatment of scale uncertainties, but also by scrutinizing a much broader range of observables. Surprisingly, $J/\psi$ hadroproduction is well described up to the highest measured values of $p_T$. Potential NRQCD based relations nontrivially lead to a perfect description of $\Upsilon(nS)$ production data. Furthermore, $J/\psi$ production in $\gamma p$ and $\gamma \gamma$ collisions is, contrary to prevailing conceptions, reproduced down to $p_T=1$~GeV, as long as the region of large inelasticity $z$ is excluded. The overall picture is much rosier than usually perceived, the more so as the remaining discrepancies appear in phase space regions where solutions via varying kinds of resummations have been proposed.}

\FullConference{QCD at the Extremes (QCDEX2025)\\
1–5 Sept 2025\\
Online\\}

\begin{document}
\maketitle

\section{Introduction and overview}

Nonrelativistic QCD~\cite{Bodwin:1994jh} (NRQCD) is arguably the most prominent and theoretically most rigorous candidate theory for quarkonium production. It conjectures a factorization of quarkonium $H$ production cross sections into nonperturbative long-distance matrix elements $\langle\mathcal{O}^H[n]\rangle$ (LDMEs) and perturbatively calculable short-distance coefficients (SDCs). Due to velocity scaling rules, the contributions from different intermediate Fock states $n$ are ordered as expansions in the relative heavy quark velocity squared $v^2$. Current quarkonium phenomenology calculates the SDCs up to next-to-leading order (NLO) in the strong coupling constant $\alpha_s$ and, for $S$ wave quarkonia, up to NLO in $v^2$. This implies for $\psi(nS)$ and $\Upsilon(nS)$ mesons $v^2$-leading $^3S_1^{[1]}$ color singlet states as well as $v^2$-subleading $^1S_0^{[8]}$, $^3S_1^{[8]}$ and $^3P_J^{[8]}$ color octet (CO) states. The CO LDMEs are obtained from fits to data, and LDMEs of different quarkonia can be related to each other via heavy quark spin symmetry~\cite{Bodwin:1994jh}, relating e.g. $J/\psi$ to $\eta_c$ LDMEs, or potential NRQCD (pNRQCD) derived relations~\cite{Brambilla:2022rjd,Brambilla:2022ayc}, relating e.g. the LDMEs of $\psi(nS)$ and $\Upsilon(nS)$. The current picture of these NLO tests of NRQCD factorization are, however, not conclusive, see e.g. the recent review~\cite{Boer:2024ylx}. Fits to $\psi(nS)$ low- and high-transverse momentum $p_T$ data~\cite{Butenschoen:2011yh,Butenschoen:2022qka} are, in particular, not able to describe polarization, $\eta_c$ data or $J/\psi+Z$ production. Fits to only high-$p_T$ data, typically $p_T>7$~GeV, can automatically describe polarization, but are known to fail to describe, in particular, lower $p_T$ hadroproduction and HERA photoproduction. These large-$p_T$ solutions are either based on a $^1S_0^{[8]}$ dominance~\cite{Bodwin:2015iua}, which leads to a disagreement with $\eta_c$ production data, or a cancellation of $^3S_1^{[8]}$ and $^3P_J^{[8]}$ CO states, where $\langle\mathcal{O}^{J/\psi}[^1S_0^{[8]}]\rangle$ can e.g. be fixed by requiring agreement with that $\eta_c$ data, as was e.g. the basis of analyses~\cite{Han:2014jya,Zhang:2014ybe}.

The latter {\em cancellation solution} is rigorously investigated and followed through in our recent analysis~\cite{Brambilla:2024iqg}. There, we apply a rigorous approach to investigate the effect of scale uncertainties, we show nontrivial agreement with a number of observables not studied in this class of fit before, and, importantly, find an understanding of the mentioned discrepancy with HERA photoproduction. In this proceedings contribution, we only want to highlight the main findings of that analysis. For all technical details, we refer directly to Ref.~\cite{Brambilla:2024iqg}.

\section{Fit results and scale uncertainty treatment\label{sec:fit}}

\begin{table}
{\centering
\begin{tabular}{c|c|c|c|c}
 & $\langle\mathcal{O}^{J/\psi}(^3S_1^{[8]})\rangle$ & $\langle\mathcal{O}^{J/\psi}(^1S_0^{[8]})\rangle$ & $\langle\mathcal{O}^{J/\psi}(^3P_0^{[8]})\rangle/m_c^2$ & $\, \, \chi^2_{\mathrm{min}}/\text{d.o.f.}\, \, $ \\
\hline 
$\, \, m_T/2\, \, $ & $\, \,0.592\pm 0.057\, \,$ & $\, \,-0.205\pm 0.196\, \,$   &   $\, \,0.697 \pm 0.089\, \,$ & $0.34$ \\\hline
$\, \,  m_T\, \, $ &  $1.050\pm 0.121$ & $ 0.068\pm 0.2489$   &  $ 1.879 \pm 0.261$ & $ 0.22$ \\ \hline
$\, \, 2m_T\, \, $& $1.382\pm 0.189$ &$0.358\pm 0.303$    &  $3.270 \pm 0.533$ & $0.21$  \\ \hline
\end{tabular}\par
}
\caption{Fit results in units of $10^{-2}$ GeV$^3$ for our three scale choices $\mu_r=\mu_f=m_T/2$, $m_T$ and $2m_T$.}
\label{tab:LDMEfit}
\end{table}

\begin{figure*}
\includegraphics[width=3.7cm]{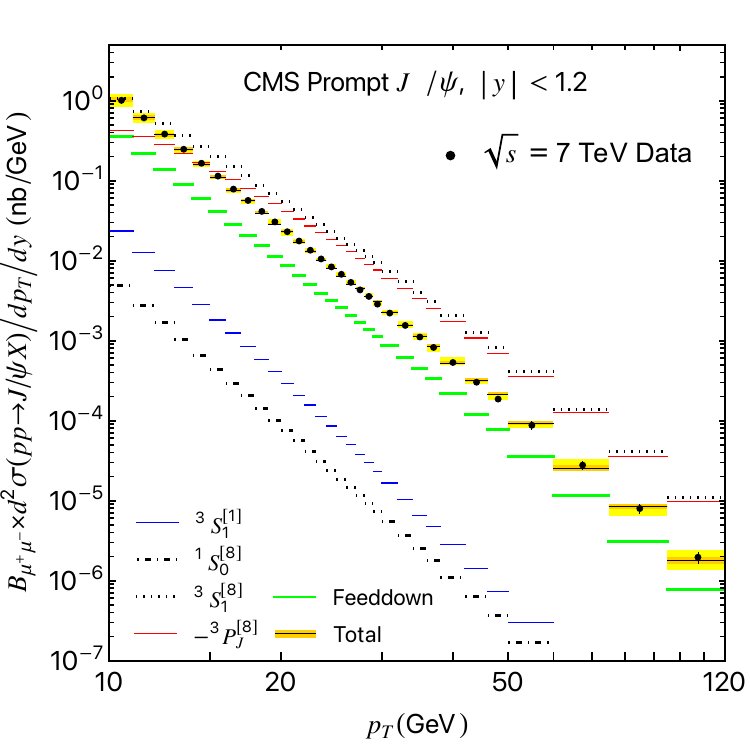}
\includegraphics[width=3.7cm]{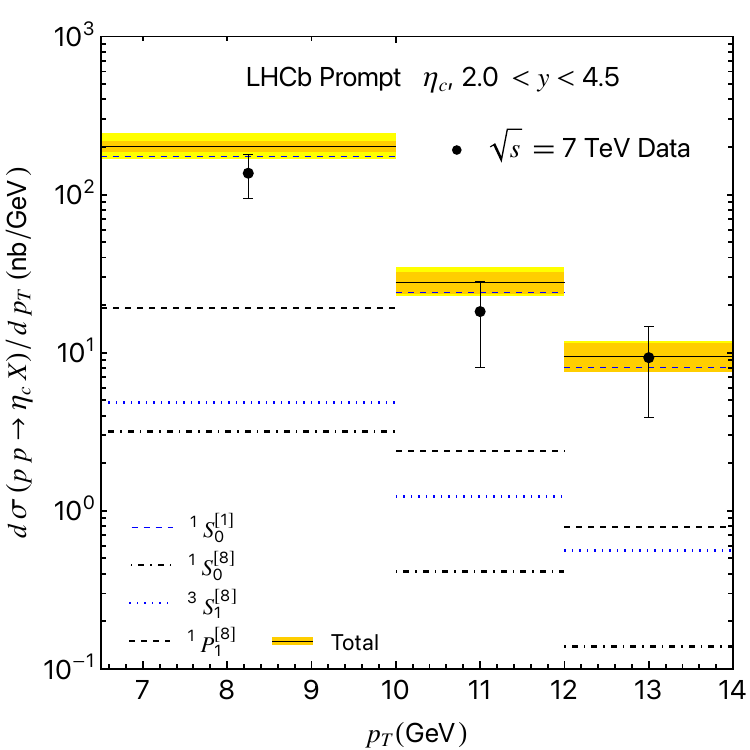}
\includegraphics[width=3.7cm]{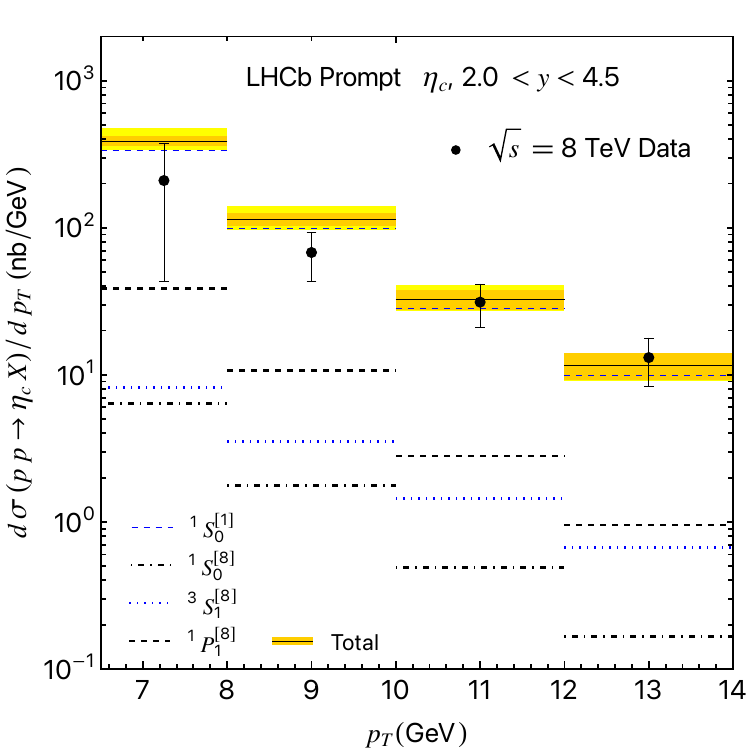}
\includegraphics[width=3.7cm]{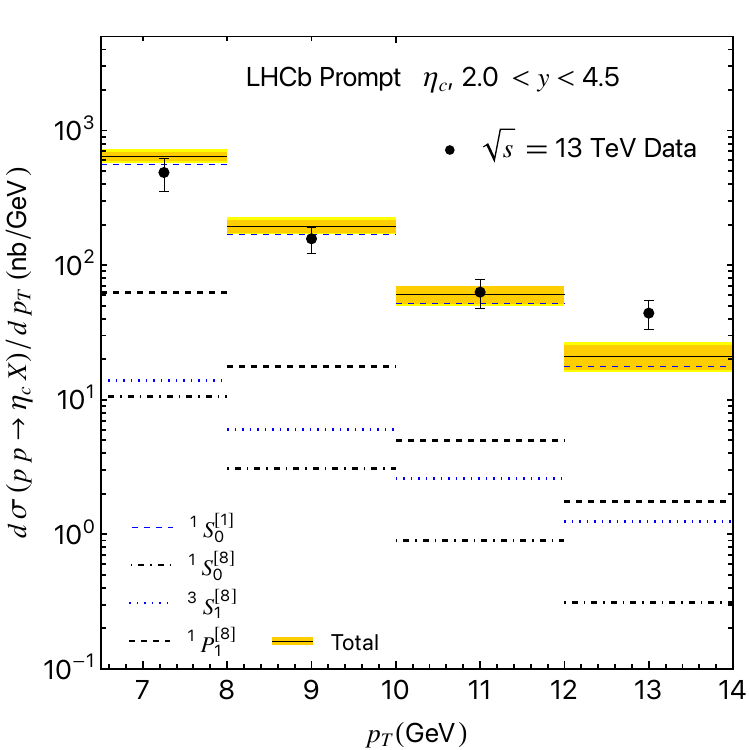}
\caption{\label{fig:fit}%
The fit to prompt $J/\psi$ production at CMS~\cite{CMS:2015lbl} and $\eta_c$ production at LHCb~\cite{LHCb:2014oii,LHCb:2024ydi}. The total cross section is broken down into feeddown contributions and the direct contributions of the individual Fock states. The orange bands describe the uncertainties from the fit correlations, the orange band additionally the effect of scale variations as described in section~\ref{sec:fit}.} 
\end{figure*}

The result of our CO LDME fits to prompt $J/\psi$ production at CMS~\cite{CMS:2015lbl} and $\eta_c$ production at LHCb~\cite{LHCb:2014oii,LHCb:2024ydi} are shown in Table~\ref{tab:LDMEfit} and Figure~\ref{fig:fit}. In Figure~\ref{fig:fit}, we recover the characteristic features of the fit, the cancellation between $^3S_1^{[8]}$ and $^3P_J^{[8]}$ states in $J/\psi$ hadroproduction on the one hand, and the constraining power of the $\eta_c$ data, which is already saturated by the color singlet contributions and pushes $\langle\mathcal{O}^{\eta_c}[^3S_1^{[8]}]\rangle\approx\langle\mathcal{O}^{J/\psi}[^1S_0^{[8]}]\rangle$ to very small values, on the other hand. We remark that the cancellation between $^3S_1^{[8]}$ and $^3P_J^{[8]}$ channels is no fine-tuning, since, at NLO in $\alpha_s$, only the sum of these contributions has physical meaning, see e.g. section 3.2 of Ref.~\cite{Butenschoen:2019lef}.

Let us now come to our treatment of the uncertainties induced by the choices of the renormalization scale $\mu_r$ and the factorization scale $\mu_f$. We perform our default fit choosing $\mu_r=\mu_f=m_T$ with $m_T$ the transverse mass  $m_T=\sqrt{p_T^2+4m_Q^2}$ and $m_Q$ the heavy quark mass. The inner orange bands of our plots are prepared with this scale choice, and show the fit uncertainties, using the full correlation matrix information of the default fit. We do, however, not know if $m_T$ is really the correct scale choice. Therefore, we perform the same fit procedure another two times, with the scale choices $2m_T$ and $m_T/2$, respectively, each resulting in different LDME fit results, different correlation matrices, different central values and shifted fit uncertainty bands. Assuming that the correct scale lies somewhere between $m_T/2$ and $2m_T$, the correct uncertainty estimate should then be the envelope of all three bands. This is the overall uncertainty we draw in yellow color  in our plots. We note that most other NRQCD analyses completely ignore the effect of scale uncertainties, or at best treat them as uncorrelated to the fit errors, a way to greatly overestimate them, in particular in a strong cancellation situation.

\section{Predictions}

\newcommand{\plotwithnumber}[2]{\begin{tikzpicture}
 \draw (0, 0) node[inner sep=0] {\includegraphics[width=3.7cm]{#1}};
 \draw (-1.55cm, -1.7cm) node {\scriptsize (#2)};
\end{tikzpicture}}

\begin{figure*}
\plotwithnumber{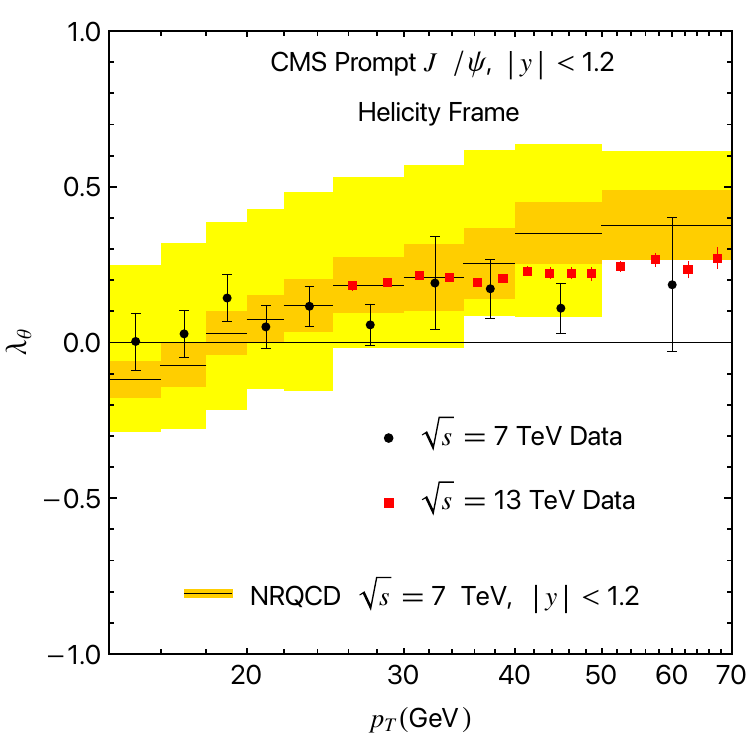}{a}
\plotwithnumber{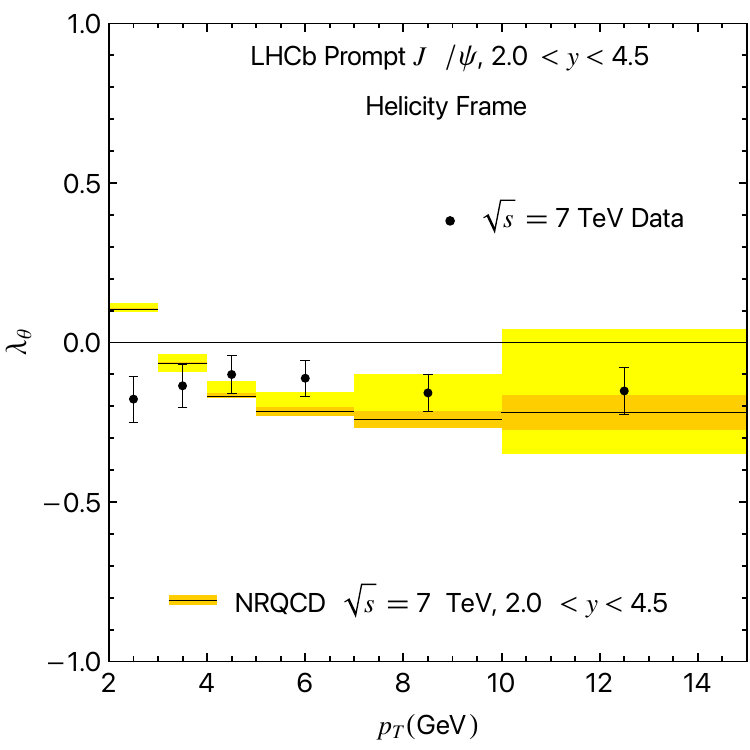}{b}
\plotwithnumber{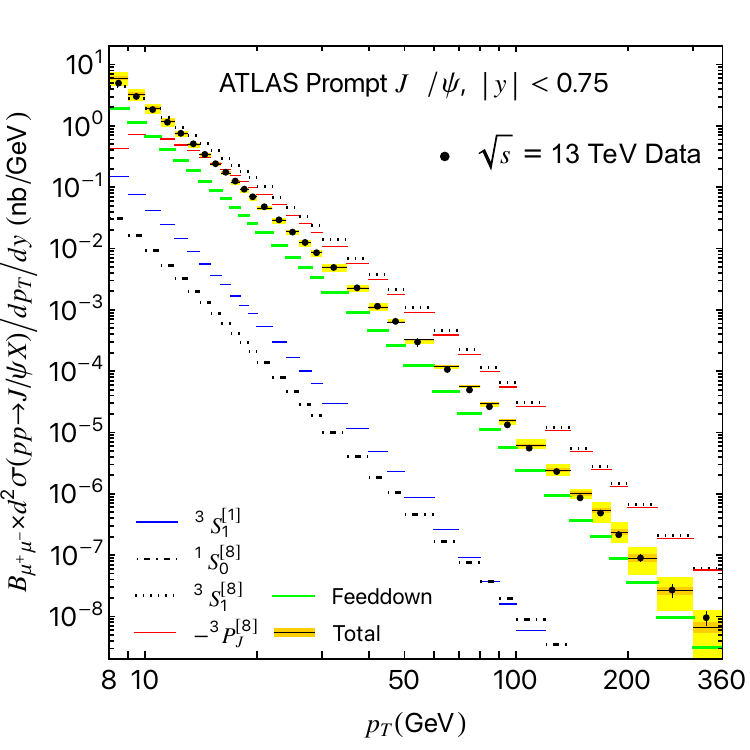}{c}
\plotwithnumber{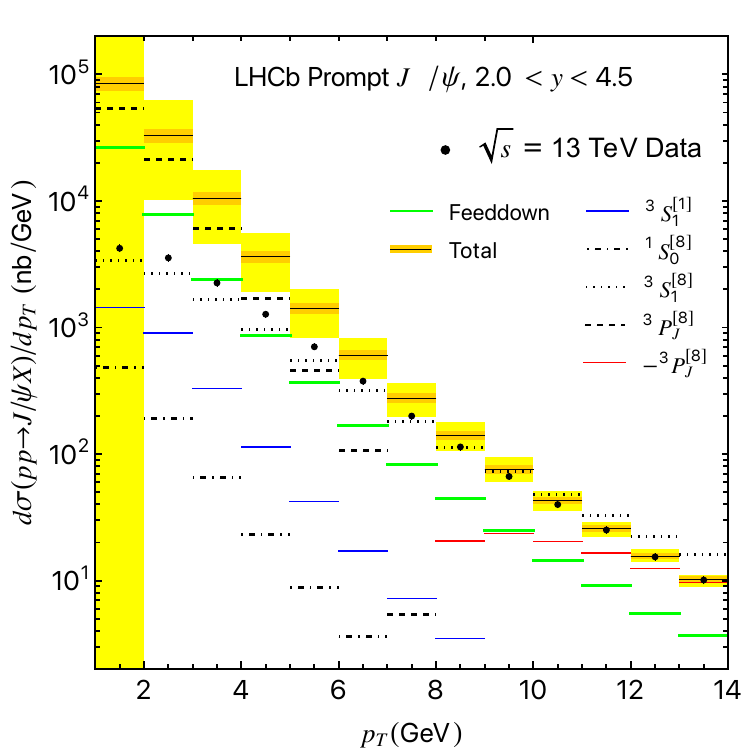}{d}
\plotwithnumber{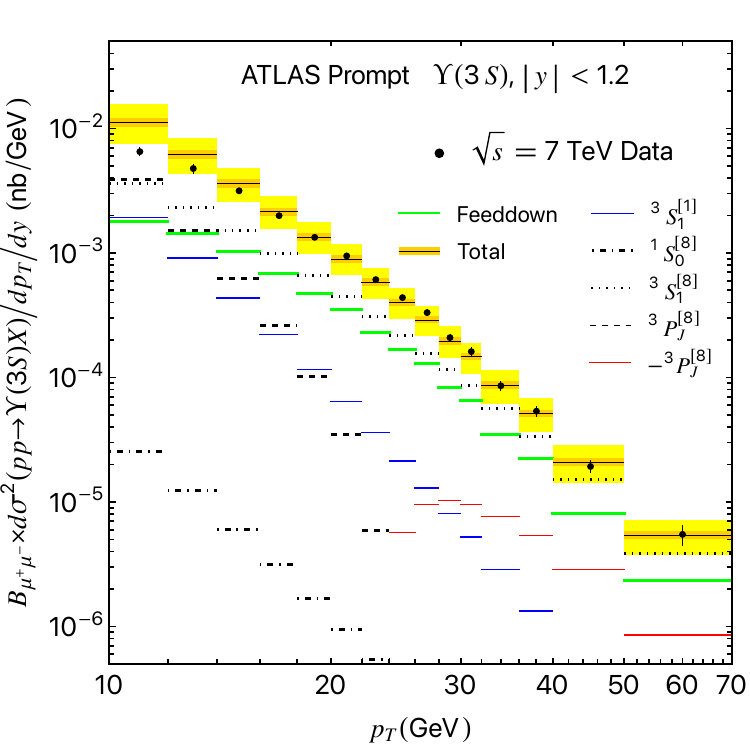}{e}
\plotwithnumber{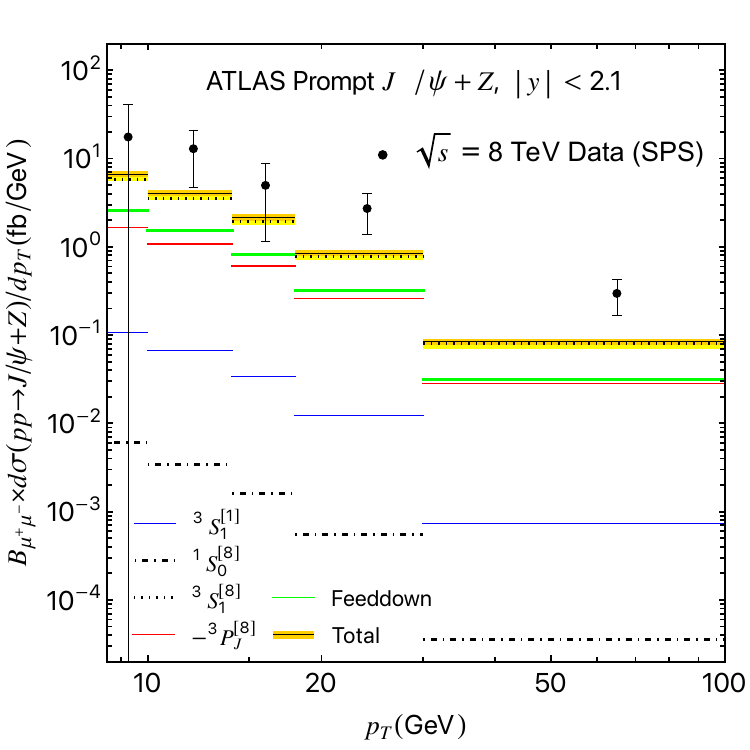}{f}
\plotwithnumber{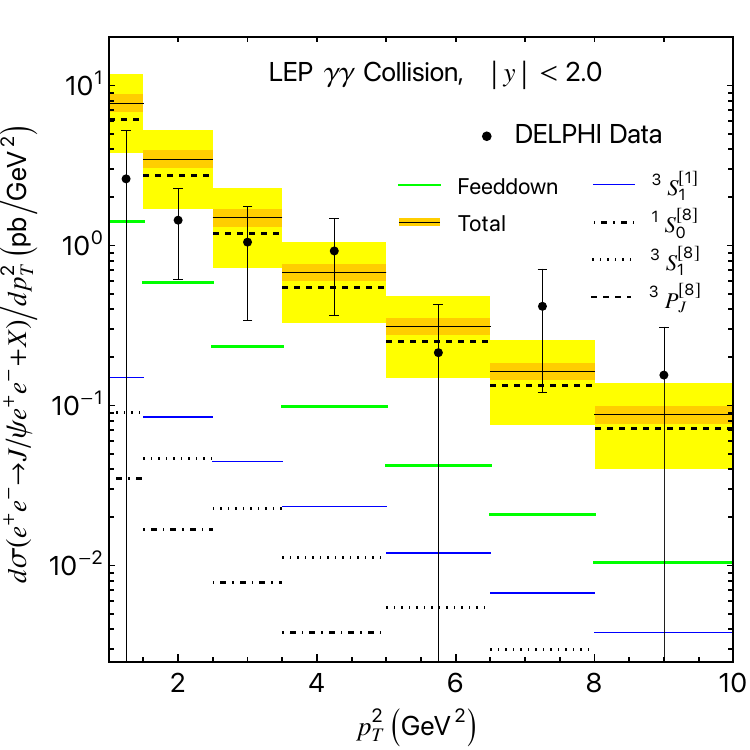}{g}
\plotwithnumber{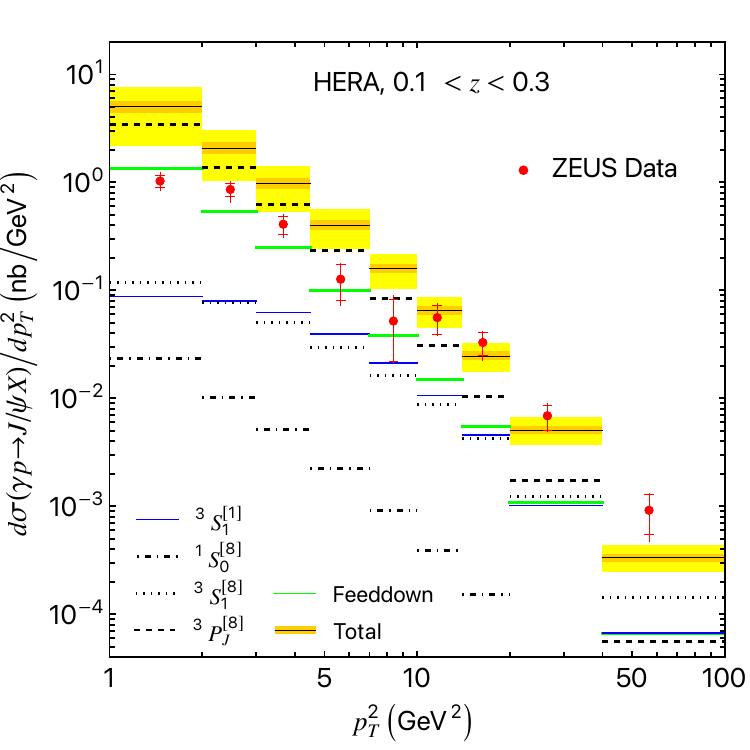}{h}
\plotwithnumber{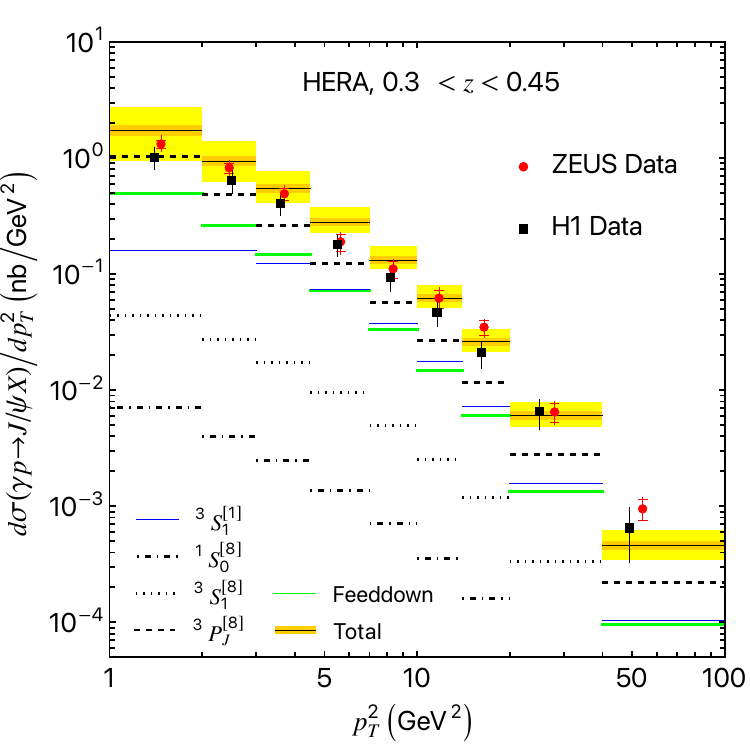}{i}
\plotwithnumber{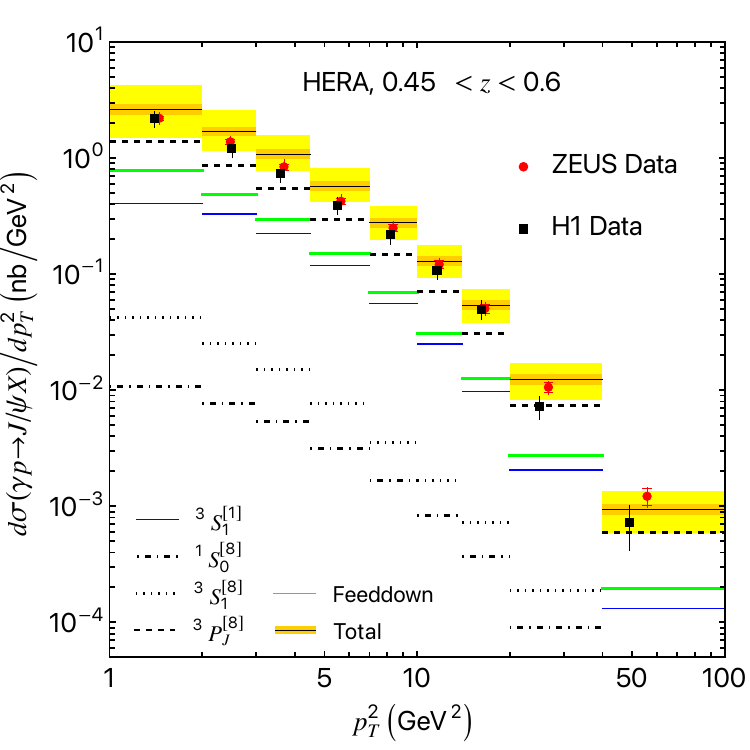}{j}
\plotwithnumber{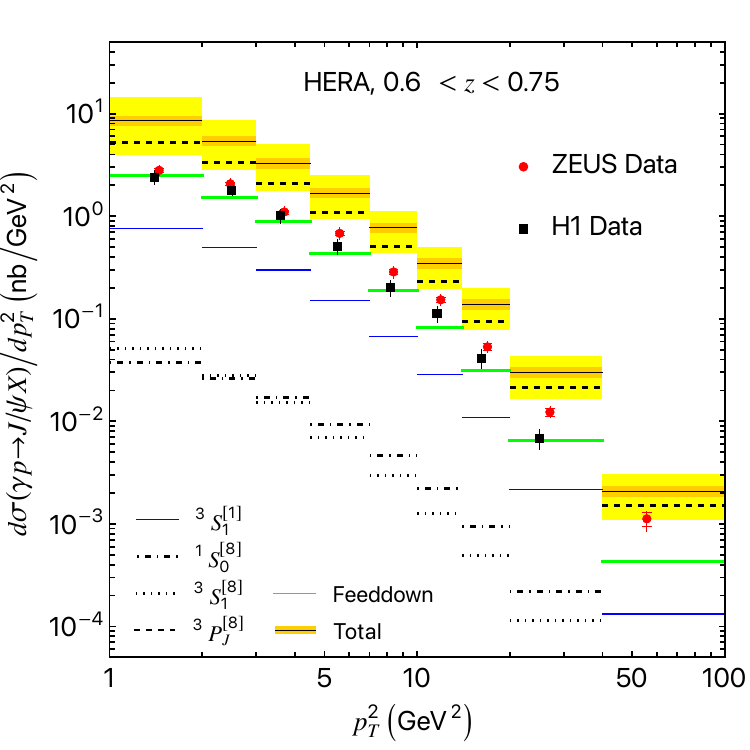}{k}
\plotwithnumber{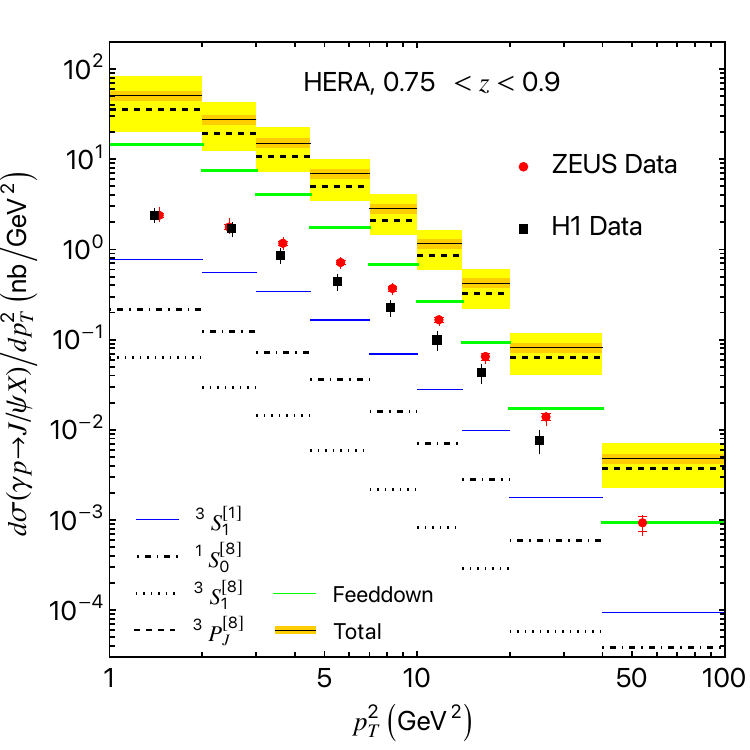}{l}
\caption{\label{fig:predictionplots}
Theory predictions for the $J/\psi$ polarization parameter $\lambda_{\theta}$ in the helicity frame compared to CMS~\cite{CMS:2013gbz,CMS:2024igk} (a) and LHCb~\cite{LHCb:2013izl} (b) data, for $J/\psi$ hadroproduction measured by ATLAS~\cite{ATLAS:2023qnh} (c) and LHCb~\cite{LHCb:2015foc} (d), for $\Upsilon(3S)$~\cite{ATLAS:2012lmu} (e) and single-parton scattering (SPS) $J/\psi+Z$~\cite{ATLAS:2014ofp} (f) production measured by ATLAS, for LEP DELPHI two photon collision data~\cite{DELPHI:2003hen} (g), and for HERA H1~\cite{H1:2010udv} and ZEUS~\cite{ZEUS:2012qog} photoproduction data (h--l). Panels (g) to (l) show the sum of non-, single- (and double-) resolved photon contributions, which are not broken down here, to avoid clutter. In all panels, the total cross sections are broken down into feeddown contributions and the direct contributions of the individual Fock states. The orange bands describe the uncertainties from the fit correlations, the yellow band additionally the effect of scale variations as described in section~\ref{sec:fit}.} 
\end{figure*}

In Figure~\ref{fig:predictionplots}, we use our LDME fit results to show predictions for a wide range of production observables. Let us paraphrase the most important findings:

\begin{itemize}
 \item In panels (a), (b) and (d), we recover the known results for  $J/\psi$ polarization in proton-proton collisions and low-$p_T$ hadroproduction yield. Helicity frame $J/\psi$ polarization is well described, largely due to the cancellation between the $^3S_1^{[8]}$ and $^3P_J^{[8]}$ transverse components. Inclusive hadroproduction below $p_T\lessapprox7$~GeV is not reproduced, due to the $^3P_J^{[8]}$ channel turning positive there, leading to an enhancement of the cross section instead of a cancellation.
 \item The excellent agreement with very-high-$p_T$ ATLAS data, up to 360~GeV, in panel (c) is unexpected, because in this region resummations of large logarithms involving $m_Q^2/p_T^2$ as well as soft-gluon resummations are expected to yield large corrections to our fixed-order results. A possible explanation could be that the resummation contributions of the $^3S_1^{[8]}$ and $^3P_J^{[8]}$ channels cancel like their NLO counterparts. This explanation is supported by Fig.~3 of Ref.~\cite{Chung:2024jfk}, where the {\em resummed} band of the lower left panel, with LDMEs similar to our default ones, describes the ATLAS data reasonably well.
 \item The prediction of $\Upsilon(3S)$ in panel (e) is achieved using pNRQCD based relations~(3.47)--(3.48) of Ref.~\cite{Brambilla:2022ayc}, which express $\Upsilon(nS)$ LDMEs in terms of $J/\psi$ LDMEs. The agreement with data is nontrivial because the three CO LDMEs evolve differently from the charm to the bottom scale, resulting in a very different Fock state decomposition than in the $J/\psi$ case, with the $^3S_1^{[8]}$ channel and feeddown from $\chi_{bJ}$ mesons contributing almost equally to the cross section. Plots for $\Upsilon(1S)$ and $\Upsilon(2S)$, with equally good agreement, were shown in the talk.
 \item The comparison to $J/\psi+Z$ production data in panel (f) works reasonably well. From the data, we have subtracted double parton scattering contributions, estimated by ATLAS via the usual pocket formula. The pocket formula itself is, however, only an approximation~\cite{Bali:2021gel}.
 \item The out-of-the-box agreement with most of the $\gamma\gamma$ scattering at LEP and HERA photoproduction in panels (g) to (l) is excellent as well. Discrepancies exist in the low-$p_T$ limit of the low inelasticity $z$ bin $0.1<z<0.3$, where the resolved photon contribution seems to import the low-$p_T$ hadroproduction problem into the photoproduction case, and for the $z\to1$ region. This result shows that it is the $z\to1$ limit, and not the low-$p_T$ region, which is the origin of the discrepancy in HERA photoproduction.
\end{itemize}

\section{Conclusions}

We have found many highly nontrival agreements with data for many observables that were not foreseen before, with the simple approach of fitting large-$p_T$ $J/\psi$ and $\eta_c$ hadroproduction data. In particular, our agreement with HERA photoproduction down to $p_T=1$~GeV for not too high $z$ casts doubt on the hypothesis that discrepancies of NRQCD predictions are due to a possible breakdown of NRQCD factorization beyond next-to-leading power in $m_Q^2/p_T^2$. The remaining regions of disagreement are low- and mid-$p_T$ hadroproduction and the $z\to1$ photoproduction region. While these discrepancies remain, we remark that they are in phase space regions where remedies have been proposed, either via small-$x$ resummations~\cite{Ma:2014mri} or via nonperturbative shape functions~\cite{Beneke:1997qw,Fleming:2006cd}.

\acknowledgments

This work was supported by the German Research Foundation through Grant No. BU~3455/1-1 as part of the Research Unit FOR2926.

\bibliographystyle{JHEP}
\bibliography{skeleton.bib}

\end{document}